\documentclass[aps,prl,reprint,groupedaddress,amsmath]{revtex4-1}

\usepackage{graphicx}
\usepackage{bm}
\usepackage{hyperref}
\usepackage{subcaption}

\newcommand{\cO}{\mathcal{O}}
\newcommand{\pder}[2]{\frac{\partial#1}{\partial#2}} 
\newcommand{\vecx}{ \textbf{x} }

\newcommand{\mean}{\overline}

\begin{document}


\title{The renormalization group in quantum quenched disorder}

\author{Vladimir Narovlansky}
\email[]{vladimir.narovlansky@weizmann.ac.il}
\author{Ofer Aharony}
\email[]{ofer.aharony@weizmann.ac.il}
\affiliation{Department of Particle Physics and Astrophysics, \\ Weizmann Institute of Science, Rehovot 7610001, Israel}

\date{\today}

\begin{abstract}
We study the renormalization group flow in general quantum field theories with quenched disorder, focusing on random quantum critical points.
We show that in disorder-averaged correlation functions the flow mixes local and non-local operators. This leads to a new crossover exponent related to the disorder (as in classical disorder).
We show that the time coordinate is rescaled at each RG step, leading to Lifshitz scaling at critical points. We write a universal formula
for the dynamical scaling exponent $z$ for weak disorder.
\end{abstract}


\maketitle

\section{Introduction}

Quantum critical points appear in many physical situations. These are scale-invariant continuous field theories, that appear at long distances and zero temperature upon fine-tuning some parameters, and that control many features of the corresponding systems with nearby parameters and with finite temperature.

Some quantum critical points depend on the existence of disorder, while others do not, 
and may (or may not) still exist when disorder is present.
We will discuss situations where the disorder is quenched (non-dynamical), so that it may be viewed as a fixed background, varying in space, for the physical system. One can then assume that there is some probability distribution for the possible disorder configurations, and compute averages of physical quantities over the disorder. In some (``self-averaging'') cases these will describe the typical behavior, while in other cases the variances may be large.

In this paper we discuss the renormalization group (RG) flow of such systems. This is particularly interesting in the vicinity of random critical points, where disorder-averaged quantities have scaling properties that are captured by critical exponents. We focus on general properties of these flows, and do not discuss any specific systems.
We discuss only zero temperature properties, though the same methods should be useful at finite temperature as well. We do not discuss various instabilities related to long-range fluctuations of the system, such as Griffiths-McCoy singularities, and it would be interesting to include them in our analysis.

Critical exponents are related to properties of local operators in the theory. We show that the RG evolution of these operators in disorder-averaged correlation functions is non-standard. Operators which are non-local in time evolve in an independent manner from the local operators, and mix with them. This leads to new critical exponents associated with the amount of disorder, analogous to the crossover exponent $\phi$ in systems with classical disorder.

We also show that the RG flow rescales time relative to space, leading to Lifshitz scaling at disordered quantum critical points, with invariance under ${\vecx}\to \vecx/b$, $t \to t / b^z $. We argue that the dynamical scaling exponent $z$ may be viewed as a (non-vanishing) beta function for a specific coupling in the action of the disordered theory, and may be computed perturbatively for weak disorder.

For some computations and for intuition we use the replica approach to disordered field theories. Perturbatively in the disorder this is just a technical trick for computing disorder-averaged quantities. Beyond perturbation theory it would be interesting to understand the implications of replica symmetry breaking for our analysis. 


A more detailed discussion of our assumptions and results may be found in the companion paper \cite{fullpaper}.

\section{The setup and the replica trick}

A particular realization of disorder is described by a disorder field $h(\vecx )$ which specifies the disorder configuration in the $d$-dimensional space parametrized by $\vecx $ (e.g., the distribution of impurities, or the strength of a background magnetic field). The disorder modifies the microscopic interactions in the system in an inhomogeneous manner. If the action of the pure system without disorder is $S_0$, in the presence of disorder the action becomes
\begin{equation} \label{eq:disorder_action}
S=S_0+ \int d^d\vecx dt\, h(\vecx )\cO _0(\vecx ,t) ,
\end{equation}
where $\cO _0$ stands for the leading interaction that disorder couples to (e.g., the order parameter when disorder is a background magnetic field). Euclidean signature is used in this entire Letter (using analytic continuation from Lorentzian). For simplicity, we assume that the disorder has spin zero, and that the pure theory is a relativistic critical system; the generalization to other situations is straightforward.

The disorder manifestly breaks space translations and space-time rotations, but we assume that the probability distribution
$P[h(\vecx)]$ to obtain a specific disorder configuration 
is translationally and rotationally invariant. The averages and all higher moments of observables are then invariant under translations and space rotations, but full relativistic invariance is not restored. 
Disorder averages are denoted by $\mean{X} \equiv \int Dh\, P[h]X$. 
We focus on short-ranged disorder correlations, such that for long distance physics the disorder range can be neglected, and the disorder at different points is statistically independent, $P[h(\vecx)] = \exp\left[-\int d^d\vecx\, p\left(h(\vecx)\right)\right]$. 
A commonly used distribution is the Gaussian one, in which $p(h)=h^2/(2v)$,
normalized to give unit sum over probabilities.

The replica trick presents quenched disorder-averages using a limit of pure field theories. For the disordered free energy we use
the identity $\log(Z) = \lim _{n \to 0} \left(\pder{Z^n}{n} \right)$, where $Z = \int D\mu \, e^{-S} $ is the partition function ($D\mu $ stands for the path integral measure in the theory). 
The replica theory is defined by
\begin{equation} \label{replica_def}
\int Dh\, P[h] Z[h] ^n= \int \prod _{A=1} ^n D\mu _A\, e^{-S_{\text{replica}}} ,
\end{equation}
having $n$ copies of the original degrees of freedom. Averaged correlation functions in the disordered theory are then related to $n \to 0$ limits of appropriate correlation functions in the replica theory.

For the Gaussian distribution 
the replica theory is
\begin{equation} \label{eq:replica_Gaussian}
S_{\text{replica}} =\sum _{A=1}^n S_{0,A} - \frac{v}{2} \sum _{A,B=1}^n \int d^d\vecx dtdt' \, \cO _{0,A} (\vecx ,t) \cO _{0,B} (\vecx ,t') .
\end{equation}
As opposed to classical disordered systems, the quantum replica theory is non-local in time, and therefore it is not clear whether we can use RG methods. 
However, this should be possible when $n \to 0$, since we have a local Wilsonian RG \cite{Osborn:1991gm} in the disordered theory.

\section{Renormalization group flow with quenched disorder}

In a pure field theory, the coupling constants $\lambda_i$ run and mix along the RG flow, and this is encoded in beta functions $\beta_i(\lambda_j)$. 
There is a one-to-one mapping between coupling constants and local operators ${\cal O}_i$, such that the action includes $\lambda_i \int {\cal O}_i$, and the local operators also mix with each other under the RG. At a fixed point $\beta_i=0$, and the derivatives of the beta functions encode the anomalous dimensions of the local operators, which are related to critical exponents.

A disordered theory is parameterized by the disorder distribution $P[h]$ in addition to the local couplings $\lambda_i$, and 
for short-range disorder this is characterized by moments $\kappa_i$
which multiply different terms in $p\left(h(\vecx)\right)$. In a specific realization of disorder the couplings $h(\vecx)$ flow, and this leads to a flow of the disorder distribution couplings $\kappa_i$. In general, the flow generates disorder for all coupling constants $\lambda_i$, with some general disorder distribution whose parameters we will still denote by $\kappa_i$. Under the RG flow, these parameters mix with the original uniform couplings $\lambda_i$, to the extent that this is allowed by the symmetries, so we have beta functions $\beta_{\lambda_i}(\lambda_j,\kappa_k)$ and $\beta_{\kappa_i}(\lambda_j,\kappa_k)$. Disorder-averaged correlation functions and thermodynamic quantities depend on all these couplings, and all of their beta functions have to vanish at disordered fixed points. Starting from a pure theory, the most relevant deformation associated with the disorder is the coupling $v$ of \eqref{eq:replica_Gaussian}, whose dimension at the fixed point is $d+2-2\Delta_0$, where $\Delta_0$ is the scaling dimension of ${\cal O}_0$ at the pure fixed point. Assuming that ${\cal O}_0$ is the lowest dimension operator allowed by the global symmetries, we have $\nu = 1 / (d+1-\Delta_0)$, so the disorder is relevant whenever $\nu < 2/d$ (this is known as the Harris criterion \cite{Harris:1974zz}).

In the next section we show that disordered theories have a special coupling constant related to rescalings of time and to the emergence of Lifshitz scaling. In the following section we then discuss the general mixing of operators in disorder-averaged theories, and show that this leads to a new critical exponent.

\section{The dynamical scaling exponent}


In this section we show that the dynamical scaling exponent $z$ behaves like an anomalous dimension of an operator -- it runs along the RG, and converges at a fixed point to the anomalous scaling exponent.
This is shown to be equivalent to a geometric emergence of Lifshitz scaling.
The discussion is general and applies also to strongly coupled theories; in fact it holds in any system (not only with disorder) breaking relativistic invariance. Without disorder the same relation of the dynamical scaling exponent to running RG couplings was discussed in \cite{Korovin:2013bua}, with some additional assumptions that do not hold in disordered systems.

Consider the replica action \eqref{eq:replica_Gaussian}.
Whenever $t'$ is close to $t$, we are allowed to use the operator product expansion (OPE), replacing the product $\cO _0 \times \cO _0$ by a series of local operators. There is one particularly interesting operator appearing universally in this OPE, which is the energy-momentum tensor $T_{\mu \nu}$.
Let $x$ stand for the space-time coordinate. Around the pure critical theory, the coefficient of this operator in the OPE is given by
\begin{equation} \label{eq:OO_T_OPE}
\cO _{0,A} (x) \cO _{0,B} (0) \supset \frac{c_{\cO \cO T} \delta_{AB}}{c_T} \frac{x^{\mu } x^{\nu }  }{x^{2\Delta _0-d+1}} T_{\mu \nu,A } (0),
\end{equation}
where $c_T$ and $c_{\cO \cO T} $ are the coefficients in the 2-point function $\langle T T \rangle$ and the 3-point function $\langle \cO_0 \cO_0 T_{\mu \nu } \rangle$, respectively \cite{fullpaper}.
Performing the integral over $t'$ in \eqref{eq:replica_Gaussian}, 
this leads to an $a$-dependent ($a$ is the lattice spacing) term proportional to $\int d^d\vecx dt\, T_{00,A} (\vecx ,t)$, which is no other than the Hamiltonian $H$ integrated over time. Along the RG flow the lattice spacing changes, and this term flows (in particular, it is generated along the RG even if it was not there to begin with). The effective action for Gaussian disorder is then given by
\begin{equation} \label{eq:Hamiltonian_deformation}
\begin{split}
& S_{\text{replica}} = \sum _{A} S_{0,A} +h_{00} \sum _{A} \int dt\, H_A - \\
&  - \frac{v}{2} \sum _{A,B} \int d^d\vecx dtdt' \, \cO _{0,A} (\vecx ,t) \cO _{0,B} (\vecx ,t')
\end{split}
\end{equation}
with a running coupling denoted by $h_{00} $. The new term
is equivalent to adding $h_{00} \int dt \, H$ to the original disordered theory. 

The integrated Hamiltonian term is actually equivalent to a stretching of time. Indeed, by Noether's theorem, under an infinitesimal transformation $x'_{\mu } =x_{\mu } +\epsilon _{\mu } $, the variation of the action is $\delta S=- \int d^d\vecx dt\, \partial _{\mu } \epsilon _{\nu } T^{\mu \nu } $. Therefore, for an infinitesimal time dilation $t'=t(1+\epsilon )$ we get precisely the new generated term $\delta S=-\epsilon \int dt\, H$.
As a result, we can either think about the RG as a flow including the coupling $h_{00} $, or alternatively we can rescale time to get rid of this term, with no $h_{00} $ coupling. For instance, for scalar operators, the correlation functions in the two approaches are related by
\begin{equation} \label{eq:relation_to_time_rescaling}
\begin{split}
& \langle \cO _{1} (x_1)\cdots \cO _{k} (x_k)\rangle_{h_{00} } =\\
& \quad = \left(1-h_{00} \sum _{i=1} ^k  t_i \pder{}{t_i} \right) \langle \cO _{1} (x_1) \cdots \cO _{k} (x_k)\rangle+O(h_{00} ^2) .
\end{split}
\end{equation}

Usually at an RG fixed point all beta functions vanish and the coupling constants flow to fixed values. However, it turns out that a constant beta function for $h_{00}$ is also consistent with scaling, and indeed generically this is what one finds at non-relativistic RG fixed points. At each RG step we rescale the cutoff (the lattice spacing), or, equivalently, we rescale the space and time coordinates by $\vecx \to \vecx / b$, $t \to t/b$. If $h_{00}$ has a constant beta function $\beta_{h_{00}}$, then under an infinitesimal RG step it changes by $h_{00} \to h_{00} - \beta_{h_{00}} \log(b)$. Naively this means that the theory is not invariant, but the discussion of the previous paragraph implies that we can equivalently keep $h_{00}$ fixed but perform an additional rescaling of the time coordinate by $t \to t (1 + \beta_{h_{00}} \log(b)) \sim t b^{\beta_{h_{00}}}$. Thus, we find that the theory is invariant under the modified scaling transformation $\vecx \to \vecx / b$, $t \to t / b^{1-\beta_{h_{00}}}$, which is a Lifshitz scaling transformation with the dynamical scaling exponent $z = 1 - \beta_{h_{00}}$. If our original pure theory has a Lifshitz scaling with a dynamical scaling exponent $z_{pure}$, then the same arguments give for the new fixed point $z_{random} = z_{pure} - \beta_{h_{00}}$.


To show how this affects correlation functions, we can look at their RG flow in the replica theory. The RG implies that when we change the RG scale $M$ and simultaneously change the coupling constants (including $v$, $h_{00} $, and other possible couplings $\lambda _i$) and allow for anomalous dimensions $\gamma$ for operators, the theory remains the same. Applied to correlation functions, this is called the Callan-Symanzik equation \cite{Callan:1970yg,Symanzik:1970rt}. The $n\to 0$ limit of the replica theory gives disorder-averaged correlation functions, and therefore we find for connected correlation functions of the lowest-dimension scalar operator $\cO $ \footnote{For other operators and for non-connected correlation functions a generalization of the Callan-Symanzik equation is needed \cite{fullpaper}.}
\begin{equation}
\begin{split}
& \bigg( M \pder{}{M} + \beta _v \pder{}{v} + \beta _{\lambda _i} \pder{}{\lambda _i} + \\
& \qquad  + \beta _{h_{00} } \pder{}{h_{00} } +k \gamma \bigg) \mean{\langle \cO (x_1) \cdots \cO (x_k)\rangle_{conn} }=0 .
\end{split}
\end{equation}
Using the relation \eqref{eq:relation_to_time_rescaling}, the 2-point function at a quantum disordered fixed point satisfies
\begin{equation} \label{rgeq}
\bigg( M \pder{}{M} +\gamma _t^* t \pder{}{t} +2 \gamma^*  \bigg) \mean{\langle \cO (x) \cO (0)\rangle_{conn} }=0 ,
\end{equation}
where $\gamma _t^* \equiv -\beta _{h_{00} } $ and $\gamma^*$ is the anomalous dimension of $\cO$ at the fixed point. Let $\Delta $ be the dimension of $\cO $ at the pure theory. The solution of \eqref{rgeq} is determined up to a function $F$ to be
\begin{equation}
\mean{ \langle \cO (x) \cO (0)\rangle_{conn} }= \frac{M^{-2\gamma ^*} }{\vecx ^{2\Delta +2\gamma ^*} } F\left( M^{\gamma ^*_t}  \frac{\vecx ^{1+\gamma _t^*} }{t }  \right) .
\end{equation}
This is indeed invariant under Lifshitz scaling $\vecx \to \vecx / b$, $t \to t / b^z$, with $\cO $ having scaling dimension $\Delta +\gamma ^*$ and with $z=1 + \gamma_t^*$ as above.
Thus, $\gamma _t^*$ plays the role of an `anomalous dimension of time'. 

For weak disorder, we can give a universal formula for the dynamical scaling exponent. We assume that the Harris criterion is saturated, $\Delta_0 = (d+2)/2$, so that $v$ is dimensionless and we can use perturbation theory.
In this case substituting \eqref{eq:OO_T_OPE} in the $v$ term of \eqref{eq:replica_Gaussian} and performing the $t'$ integration gives the singular in $a$ term
\begin{equation}
\frac{v c_{\cO \cO T} }{c_T} \log(a)\sum _A \int dt \, H_A .
\end{equation}
The flow of $h_{00} $ should be such as to compensate for this lattice spacing dependence, giving the beta function $ \beta _{h_{00} } = v c_{\cO \cO T}/c_T + O(v^2)$.
If we normalize the two-point function of $\cO_0$ to one, and use the conformal Ward identity to compute $c_{\cO \cO T}$, we obtain at leading order in $v$
\begin{equation} \label{eq:perturbative_z}
z \approx 1+ \frac{v}{2 c_T} 
\frac{(d+1)(d+2)}{d} \frac{\Gamma \left( (d+1)/2 \right)}{2\pi ^{(d+1)/2} } .
\end{equation}
This formula is valid for any theory with weak (marginal) disorder. In particular, it reproduces the strongly coupled holographic result in \cite{Hartnoll:2014cua} and the weakly coupled result in \cite{Boyanovsky:1982zz}. In the former, $z$ was computed using holography, and here we see the field theory interpretation of this, with the same numerical value. 

Note that \eqref{eq:perturbative_z} used \eqref{eq:replica_Gaussian} in which Gaussian disorder is assumed. However, it is still true for a generic disorder distribution with variance $v$, since for marginal disorder the corrections to \eqref{eq:replica_Gaussian} from higher disorder moments are irrelevant.

\section{Operator mixings and disorder critical exponents}

For classical disorder, the replica theory \eqref{replica_def} is local, and the moments of the disorder distribution are standard coupling constants of the replica theory. Thus 
the disordered RG flow is an $n\to 0$ limit of standard RG flows, with general mixings of all couplings $\{\lambda_i, \kappa_i\}$ \cite{Emery:1975zz,Grinstein:1976zz}. In order to flow to a (disordered) fixed point, any coupling related to the disorder must be irrelevant at the fixed point. Since the couplings flow independently and mix, the associated critical exponent is no longer directly related to $\Delta_0$ (and to $\nu$) as it was at short distances, but rather there is a new crossover exponent $\phi$, determined by the dimension of the leading coupling related to the disorder.

For quantum disorder the situation is different, since the disorder couplings multiply non-local operators in the replica theory. For Gaussian disorder there are two integrations over the time direction \eqref{eq:replica_Gaussian}, while $k$'th moments of the disorder distribution multiply terms with $k$ integrations over the time direction \footnote{When $\Delta_0 \leq 1$, this implies that there is an infinite number of relevant operators near the pure fixed point, and the flow of the full disorder distribution must be considered; we will assume here that $\Delta_0 > 1$.}. 
Naively, since in such non-local operators the different ${\cal O}_0$'s are separated in time, one may expect their renormalization to be determined by that of the local operators ${\cal O}_0(\vecx,t)$. 
So one may guess that the mixings of $\lambda_i$ and $\kappa_i$ described above do not occur, and that the scaling dimensions of non-local terms in the action like \eqref{eq:replica_Gaussian} are determined by those of the local operators ${\cal O}_0(\vecx,t)$; there would then be no independent critical exponent $\phi$ associated with the disorder. We will show that these expectations are actually not correct. Note that the non-locality leads to non-renormalizable divergences for $n\neq 0$, but the theory becomes renormalizable as $n\to 0$.

In fact, we claim that in disorder-averaged correlation functions, general multi-local operators of the form
\begin{equation} \label{non_local_ops}
{\cal O}_1(\vecx, t) \int dt_2 {\cal O}_2(\vecx, t_2) \cdots \int dt_k {\cal O}_k(\vecx, t_k)
\end{equation}
with different values of $k$ mix with each other. The integrals of these operators multiply the $\lambda_i$ and the $\kappa_j$, so this gives rise to generic mixings of all these couplings. From the point of view of the replica theory, where disorder-averaged correlation functions of these operators are related to those of
\begin{equation} \label{non_local_ops_replica}
{\cal O}_{1,A} (\vecx, t) \sum_{A_2,\cdots,A_k=1}^n \int dt_2 {\cal O}_{2,A_2}(\vecx, t_2) \cdots \int dt_k {\cal O}_{k,A_k}(\vecx, t_k),
\end{equation}
such mixings appear naturally. When two operators in \eqref{non_local_ops_replica} from the same replica ${\cal O}_{i,A}(\vecx,t_i)$ and ${\cal O}_{j,A}(\vecx,t_j)$ approach each other in time, there is a short-distance singularity, and regularizing it leads to a mixing with another operator ${\cal O}_{k,A}(\vecx,t_i)$; the resulting operator has one fewer time integration than the original operator  \eqref{non_local_ops_replica}. We already saw an example of this in the previous section. Conversely, the perturbative-in-disorder corrections to a local operator ${\cal O}_{k,A}(\vecx,t_0)$ can be described by bringing down a disorder interaction \eqref{eq:replica_Gaussian} from the action. There is then a singularity when (say) $t$ approaches $t_0$ for arbitrary $t'$, and regularizing it requires mixing this operator with ${\cal O}_{j,A}(\vecx,t_0) \sum_{B=1}^n \int dt' {\cal }O_{0,B}(\vecx,t')$.

From the point of view of the disordered theory, such mixings seem very surprising, since this theory is local, and operators at different times cannot mix. This is true, but in disorder-averaged correlation functions, the fact that the disorder distribution is independent of time reproduces this mixing effect, whenever the mixing of local operators depends on the disordered couplings $h(\vecx)$. For instance, consider a situation where some operator ${\cal O}_i(\vecx,t)$ mixes under the RG flow in a specific realization of disorder with $h(\vecx) {\cal O}_j(\vecx,t)$, and suppose that the disorder distribution is Gaussian. In such a situation a disorder-averaged correlation function of ${\cal O}_i$,
\begin{equation}
\overline{\langle {\cal O}_i(\vecx,t) \cdots \rangle} =
\int Dh P[h] \langle {\cal O}_i(\vecx,t) \cdots \rangle_{h(\vecx)}
\end{equation}
mixes with
\begin{equation}
\begin{split}
& \int Dh P[h] h(\vecx) \langle {\cal O}_j(\vecx,t) \cdots \rangle_{h(\vecx)} = \\
& \qquad= \int Dh (-v \frac{\delta P[h]}{\delta h(\vecx)})  \langle {\cal O}_j(\vecx,t) \cdots \rangle_{h(\vecx)} \\
& \qquad= v \int Dh P[h] \frac{\delta}{\delta h(\vecx)} \langle {\cal O}_j(\vecx,t) \cdots \rangle_{h(\vecx)}.
\end{split}
\end{equation}
But the derivative with respect to $h(\vecx)$ brings down from the path integral of the disordered theory an operator $\int dt' {\cal O}_0(\vecx,t')$.
So we find that correlation functions of $ {\cal O}_i(\vecx,t)$ mix with those of ${\cal O}_j(\vecx,t) \int dt' {\cal O}_0(\vecx,t')$. One can show that such arguments account for all the non-local mixings seen in the replica theory, so these are not just artifacts of the replica description.

A specific implication of this RG analysis is that the renormalization of the non-local replica operators related to the $\kappa_i$ is independent of that of the local operators. In particular, the dimension of the disorder operator multiplying $v$ in \eqref{eq:replica_Gaussian} is not directly related to that of ${\cal O}_0$, but rather gives rise at a fixed point to an independent crossover exponent $\phi$, governing the effect of disorder at and near that fixed point. We will give an explicit example of this below. It would be interesting to measure this critical exponent at disordered quantum critical points. In order to flow to the disordered fixed point, this disordered coupling must be irrelevant, but it is not directly related to $\nu$, so from this point of view the relation $\nu > 2/d$ is not required to hold at a disordered fixed point. Nevertheless, there are independent arguments that this relation must hold \cite{Chayes:1986ju}.


\section{An example}

We give a simple example showing explicitly that the disorder operator (multiplying $v$ in \eqref{eq:replica_Gaussian}) has an independent scaling dimension, rather than being twice the dimension of $\cO _0$. The model we use is a simple variant of the one used in \cite{Boyanovsky:1982zz}; a real scalar field $\varphi $ with disorder coupled to $\varphi ^2$, related to the random-bond Ising model. But we study this model for $4$ space dimensions, where the disorder saturates the Harris bound. The replica action is
\begin{equation}
\begin{split}
& S_{replica} = \frac{1}{2}  \sum _{A=1}^n \int d^4\vecx dt\,  \left[ \sum _{i=1}^4 (\partial _i \varphi_A)^2+ \alpha (\partial _t \varphi_A)^2\right] - \\
& \qquad - \frac{v}{2} \sum _{A,B=1}^n \int d^4\vecx dt dt' \, \varphi _A(\vecx ,t)^2 \varphi _B(\vecx ,t')^2 .
\end{split}
\end{equation}
The need for the running coupling $\alpha $ in this case was noticed in \cite{Boyanovsky:1982zz}. In fact it is a special case of our general discussion above, since the Hamiltonian deformation in \eqref{eq:Hamiltonian_deformation} reduces to this action \cite{fullpaper} with $\alpha = 2 h_{00} + 1$.
The computations below are performed with this action in the $n\to 0$ limit.

A standard field theory computation gives the beta function of $v$ and its dimension
$[v]= 4v/\pi ^2+O(v^2)$. Therefore, the dimension of the operator $\Psi(\vecx ) \equiv \sum_{A,B} \int dt dt' \, \varphi _A(\vecx ,t)^2 \varphi _B(\vecx ,t')^2$ has to be $[\Psi]=4-4v/\pi ^2+O(v^2)$. On the other hand,
computing the dimension of $\varphi ^2$ gives $[\varphi ^2]=3-v/(2\pi ^2)+O(v^2)$. We see that subtracting the dimensions of the time integrals from $2[\varphi ^2]$ does not reproduce $[\Psi ]$.

\begin{figure}[t!]
    \begin{subfigure}[t]{0.15\textwidth}
          \includegraphics[height=0.6\textwidth]{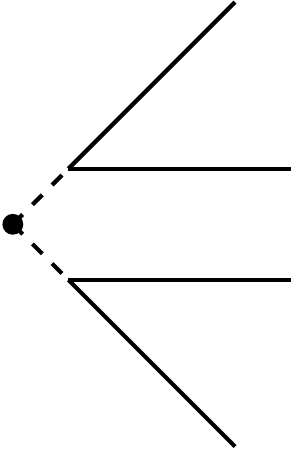}
          \caption{}
          \label{fig:Lifshitz_5d_disorder_op_dim_1}
    \end{subfigure}
    \begin{subfigure}[t]{0.15\textwidth}
          \includegraphics[height=0.6\textwidth]{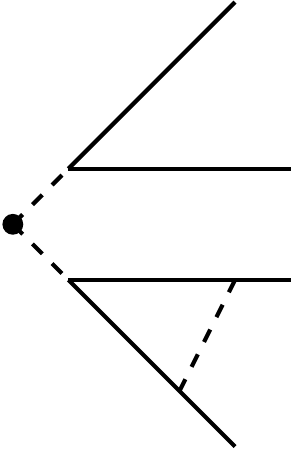}
          \caption{}
          \label{fig:Lifshitz_5d_disorder_op_dim_2}
    \end{subfigure}
    \begin{subfigure}[t]{0.15\textwidth}
          \includegraphics[height=0.6\textwidth]{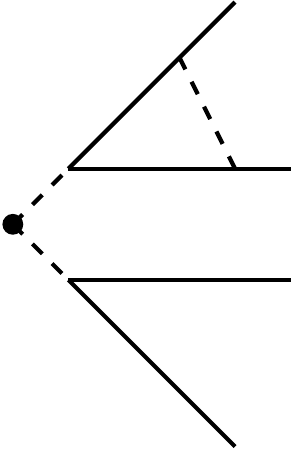}
          \caption{}
          \label{fig:Lifshitz_5d_disorder_op_dim_3}
    \end{subfigure}
    \begin{subfigure}[t]{0.15\textwidth}
          \includegraphics[height=0.6\textwidth]{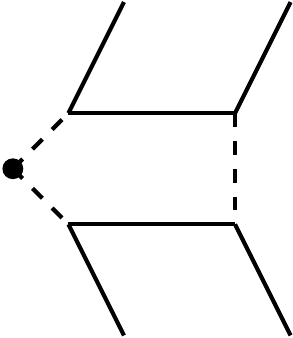}
          \caption{}
          \label{fig:Lifshitz_5d_disorder_op_dim_4}
    \end{subfigure}
    \begin{subfigure}[t]{0.15\textwidth}
          \includegraphics[height=0.6\textwidth]{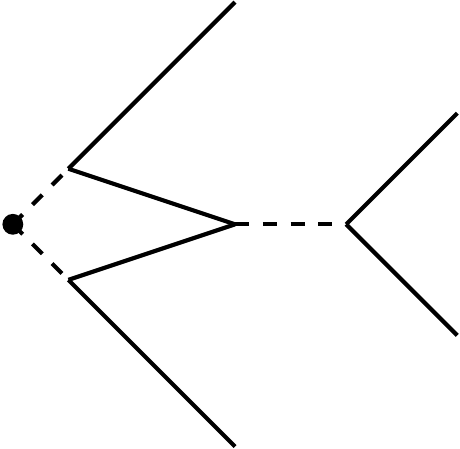}
          \caption{}
          \label{fig:Lifshitz_5d_disorder_op_dim_5}
    \end{subfigure}
    \caption{The diagrams contributing to the anomalous dimension of the (non-local in time) disorder operator $\Psi$. The solid lines are propagators, and the dashed lines correspond to the non-local $v$ interaction. The dot in each diagram, with 2 dashed lines attached, represents $\Psi(\vecx)$.}
\end{figure}

Instead, we should consider $\Psi $ as an independent operator. We can 
alternatively compute its dimension by
considering the correlation function $\langle \Psi (\vecx ) \sum _{A_1} \varphi _{A_1} (\vecx _1,t_1) \cdots \sum _{A_4} \varphi _{A_4}(\vecx _4,t_4)\rangle$. In Fig.\ \ref{fig:Lifshitz_5d_disorder_op_dim_1} we show the tree level contribution to this correlation function.
At the next order we find the corrections of Figs.\ \ref{fig:Lifshitz_5d_disorder_op_dim_2} and \ref{fig:Lifshitz_5d_disorder_op_dim_3}, which correspond to the anomalous dimensions of each of the $\varphi ^2$ factors in $\Psi $. The naive claim that the dimension of $\Psi $ is fixed by $[\varphi ^2]$ corresponds to taking into account only these corrections. However, there are additional corrections shown in Figs.\ \ref{fig:Lifshitz_5d_disorder_op_dim_4} and \ref{fig:Lifshitz_5d_disorder_op_dim_5} (there are no other diagrams contributing to the anomalous dimension as $n \to 0$). 
The sum of these corrections gives precisely the value of $[\Psi ]$ above.

\begin{acknowledgments}
We would like to thank A.~Aharony, E.~Altman, M.~Berkooz, S.~Hartnoll, Z.~Komargodski, Y.~Korovin, S.~Yankielowicz and P.~Young for useful discussions.
This work was supported in part  by the I-CORE program of the Planning and Budgeting Committee and the Israel Science Foundation (grant number 1937/12), by an Israel Science Foundation center for excellence grant, by the Minerva foundation with funding from the Federal German Ministry for Education and Research, and by the ISF within the ISF-UGC joint research program framework (grant no.\ 1200/14). OA is the Samuel Sebba Professorial Chair of Pure and Applied Physics.  
\end{acknowledgments}

\bibliography{disorder}

\begin{thebibliography}{13}%
\makeatletter
\providecommand \@ifxundefined [1]{%
 \@ifx{#1\undefined}
}%
\providecommand \@ifnum [1]{%
 \ifnum #1\expandafter \@firstoftwo
 \else \expandafter \@secondoftwo
 \fi
}%
\providecommand \@ifx [1]{%
 \ifx #1\expandafter \@firstoftwo
 \else \expandafter \@secondoftwo
 \fi
}%
\providecommand \natexlab [1]{#1}%
\providecommand \enquote  [1]{``#1''}%
\providecommand \bibnamefont  [1]{#1}%
\providecommand \bibfnamefont [1]{#1}%
\providecommand \citenamefont [1]{#1}%
\providecommand \href@noop [0]{\@secondoftwo}%
\providecommand \href [0]{\begingroup \@sanitize@url \@href}%
\providecommand \@href[1]{\@@startlink{#1}\@@href}%
\providecommand \@@href[1]{\endgroup#1\@@endlink}%
\providecommand \@sanitize@url [0]{\catcode `\\12\catcode `\$12\catcode
  `\&12\catcode `\#12\catcode `\^12\catcode `\_12\catcode `\%12\relax}%
\providecommand \@@startlink[1]{}%
\providecommand \@@endlink[0]{}%
\providecommand \url  [0]{\begingroup\@sanitize@url \@url }%
\providecommand \@url [1]{\endgroup\@href {#1}{\urlprefix }}%
\providecommand \urlprefix  [0]{URL }%
\providecommand \Eprint [0]{\href }%
\providecommand \doibase [0]{http://dx.doi.org/}%
\providecommand \selectlanguage [0]{\@gobble}%
\providecommand \bibinfo  [0]{\@secondoftwo}%
\providecommand \bibfield  [0]{\@secondoftwo}%
\providecommand \translation [1]{[#1]}%
\providecommand \BibitemOpen [0]{}%
\providecommand \bibitemStop [0]{}%
\providecommand \bibitemNoStop [0]{.\EOS\space}%
\providecommand \EOS [0]{\spacefactor3000\relax}%
\providecommand \BibitemShut  [1]{\csname bibitem#1\endcsname}%
\let\auto@bib@innerbib\@empty
\bibitem [{\citenamefont {Aharony}\ and\ \citenamefont
  {Narovlansky}(2018)}]{fullpaper}%
  \BibitemOpen
  \bibfield  {author} {\bibinfo {author} {\bibfnamefont {O.}~\bibnamefont
  {Aharony}}\ and\ \bibinfo {author} {\bibfnamefont {V.}~\bibnamefont
  {Narovlansky}},\ }\href@noop {} {\  (\bibinfo {year} {2018})},\ \Eprint
  {http://arxiv.org/abs/1803.08534} {arXiv:1803.08534 [hep-th]} \BibitemShut
  {NoStop}%
\bibitem [{\citenamefont {Osborn}(1991)}]{Osborn:1991gm}%
  \BibitemOpen
  \bibfield  {author} {\bibinfo {author} {\bibfnamefont {H.}~\bibnamefont
  {Osborn}},\ }\href {\doibase 10.1016/0550-3213(91)80030-P} {\bibfield
  {journal} {\bibinfo  {journal} {Nucl. Phys.}\ }\textbf {\bibinfo {volume}
  {B363}},\ \bibinfo {pages} {486} (\bibinfo {year} {1991})}\BibitemShut
  {NoStop}%
\bibitem [{\citenamefont {Harris}(1974)}]{Harris:1974zz}%
  \BibitemOpen
  \bibfield  {author} {\bibinfo {author} {\bibfnamefont {A.~B.}\ \bibnamefont
  {Harris}},\ }\href@noop {} {\bibfield  {journal} {\bibinfo  {journal} {J.
  Phys. C}\ }\textbf {\bibinfo {volume} {7}},\ \bibinfo {pages} {1671}
  (\bibinfo {year} {1974})}\BibitemShut {NoStop}%
\bibitem [{\citenamefont {Korovin}\ \emph {et~al.}(2013)\citenamefont
  {Korovin}, \citenamefont {Skenderis},\ and\ \citenamefont
  {Taylor}}]{Korovin:2013bua}%
  \BibitemOpen
  \bibfield  {author} {\bibinfo {author} {\bibfnamefont {Y.}~\bibnamefont
  {Korovin}}, \bibinfo {author} {\bibfnamefont {K.}~\bibnamefont {Skenderis}},
  \ and\ \bibinfo {author} {\bibfnamefont {M.}~\bibnamefont {Taylor}},\ }\href
  {\doibase 10.1007/JHEP08(2013)026} {\bibfield  {journal} {\bibinfo  {journal}
  {JHEP}\ }\textbf {\bibinfo {volume} {08}},\ \bibinfo {pages} {026} (\bibinfo
  {year} {2013})},\ \Eprint {http://arxiv.org/abs/1304.7776} {arXiv:1304.7776
  [hep-th]} \BibitemShut {NoStop}%
\bibitem [{\citenamefont {Callan}(1970)}]{Callan:1970yg}%
  \BibitemOpen
  \bibfield  {author} {\bibinfo {author} {\bibfnamefont {C.~G.}\ \bibnamefont
  {Callan}, \bibfnamefont {Jr.}},\ }\href {\doibase 10.1103/PhysRevD.2.1541}
  {\bibfield  {journal} {\bibinfo  {journal} {Phys. Rev.}\ }\textbf {\bibinfo
  {volume} {D2}},\ \bibinfo {pages} {1541} (\bibinfo {year}
  {1970})}\BibitemShut {NoStop}%
\bibitem [{\citenamefont {Symanzik}(1970)}]{Symanzik:1970rt}%
  \BibitemOpen
  \bibfield  {author} {\bibinfo {author} {\bibfnamefont {K.}~\bibnamefont
  {Symanzik}},\ }\href {\doibase 10.1007/BF01649434} {\bibfield  {journal}
  {\bibinfo  {journal} {Commun. Math. Phys.}\ }\textbf {\bibinfo {volume}
  {18}},\ \bibinfo {pages} {227} (\bibinfo {year} {1970})}\BibitemShut
  {NoStop}%
\bibitem [{Note1()}]{Note1}%
  \BibitemOpen
  \bibinfo {note} {For other operators and for non-connected correlation
  functions a generalization of the Callan-Symanzik equation is needed \cite
  {fullpaper}.}\BibitemShut {Stop}%
\bibitem [{\citenamefont {Hartnoll}\ and\ \citenamefont
  {Santos}(2014)}]{Hartnoll:2014cua}%
  \BibitemOpen
  \bibfield  {author} {\bibinfo {author} {\bibfnamefont {S.~A.}\ \bibnamefont
  {Hartnoll}}\ and\ \bibinfo {author} {\bibfnamefont {J.~E.}\ \bibnamefont
  {Santos}},\ }\href {\doibase 10.1103/PhysRevLett.112.231601} {\bibfield
  {journal} {\bibinfo  {journal} {Phys. Rev. Lett.}\ }\textbf {\bibinfo
  {volume} {112}},\ \bibinfo {pages} {231601} (\bibinfo {year} {2014})},\
  \Eprint {http://arxiv.org/abs/1402.0872} {arXiv:1402.0872 [hep-th]}
  \BibitemShut {NoStop}%
\bibitem [{\citenamefont {Boyanovsky}\ and\ \citenamefont
  {Cardy}(1982)}]{Boyanovsky:1982zz}%
  \BibitemOpen
  \bibfield  {author} {\bibinfo {author} {\bibfnamefont {D.}~\bibnamefont
  {Boyanovsky}}\ and\ \bibinfo {author} {\bibfnamefont {J.~L.}\ \bibnamefont
  {Cardy}},\ }\href {\doibase 10.1103/PhysRevB.26.154} {\bibfield  {journal}
  {\bibinfo  {journal} {Phys. Rev.}\ }\textbf {\bibinfo {volume} {B26}},\
  \bibinfo {pages} {154} (\bibinfo {year} {1982})}\BibitemShut {NoStop}%
\bibitem [{\citenamefont {Emery}(1975)}]{Emery:1975zz}%
  \BibitemOpen
  \bibfield  {author} {\bibinfo {author} {\bibfnamefont {V.~J.}\ \bibnamefont
  {Emery}},\ }\href {\doibase 10.1103/PhysRevB.11.239} {\bibfield  {journal}
  {\bibinfo  {journal} {Phys. Rev.}\ }\textbf {\bibinfo {volume} {B11}},\
  \bibinfo {pages} {239} (\bibinfo {year} {1975})}\BibitemShut {NoStop}%
\bibitem [{\citenamefont {Grinstein}\ and\ \citenamefont
  {Luther}(1976)}]{Grinstein:1976zz}%
  \BibitemOpen
  \bibfield  {author} {\bibinfo {author} {\bibfnamefont {G.}~\bibnamefont
  {Grinstein}}\ and\ \bibinfo {author} {\bibfnamefont {A.}~\bibnamefont
  {Luther}},\ }\href {\doibase 10.1103/PhysRevB.13.1329} {\bibfield  {journal}
  {\bibinfo  {journal} {Phys. Rev.}\ }\textbf {\bibinfo {volume} {B13}},\
  \bibinfo {pages} {1329} (\bibinfo {year} {1976})}\BibitemShut {NoStop}%
\bibitem [{Note2()}]{Note2}%
  \BibitemOpen
  \bibinfo {note} {When $\Delta _0 \leq 1$, this implies that there is an
  infinite number of relevant operators near the pure fixed point, and the flow
  of the full disorder distribution must be considered; we will assume here
  that $\Delta _0 > 1$.}\BibitemShut {Stop}%
\bibitem [{\citenamefont {Chayes}\ \emph {et~al.}(1986)\citenamefont {Chayes},
  \citenamefont {Chayes}, \citenamefont {Fisher},\ and\ \citenamefont
  {Spencer}}]{Chayes:1986ju}%
  \BibitemOpen
  \bibfield  {author} {\bibinfo {author} {\bibfnamefont {J.~T.}\ \bibnamefont
  {Chayes}}, \bibinfo {author} {\bibfnamefont {L.}~\bibnamefont {Chayes}},
  \bibinfo {author} {\bibfnamefont {D.~S.}\ \bibnamefont {Fisher}}, \ and\
  \bibinfo {author} {\bibfnamefont {T.}~\bibnamefont {Spencer}},\ }\href
  {\doibase 10.1103/PhysRevLett.57.2999} {\bibfield  {journal} {\bibinfo
  {journal} {Phys. Rev. Lett.}\ }\textbf {\bibinfo {volume} {57}},\ \bibinfo
  {pages} {2999} (\bibinfo {year} {1986})}\BibitemShut {NoStop}%
\end{thebibliography}%

\end{document}